\documentclass[a4paper]{elsarticle}

\usepackage{lineno}
\modulolinenumbers[5]
\usepackage{textcomp}

\usepackage[normalem]{ulem}
\useunder{\uline}{\ul}{}
\usepackage{placeins}
\usepackage{amssymb,mathtools}
\usepackage{url}
\usepackage{adjustbox}
\usepackage[nottoc]{tocbibind}
\usepackage{color}

\usepackage{xcolor}

\journal{ArXiv}

\begin{document}

\begin{frontmatter}

\title{Feasibility of Equity-driven Taxi Pricing Strategy based on Double Auction Mechanism in Bangkok Metropolitan Region, Thailand}

\author[1]{He-in Cheong}
\affiliation[1]{organization={Imperial College London},
            addressline={Skempton Building, South Kensington Campus}, 
            city={London},
            postcode=\mbox{{SW7 2BU}}, 
            state={London},
            country={United Kingdom}}
            
\author[2]{Jonathan Sanz Carcelen}
\affiliation[2]{organization={Universitat Politecnica de Catalunya},
            addressline={Campus Nord, Carrer de Jordi Girona, 1,3}, 
            city={Barcelona},
            postcode={08034}, 
            state={Catalonia},
            country={Spain}}

\author[1]{Manlika Sukitpaneenit}
\author[1]{Panagiotis Angeloudis}
\author[1]{Arnab Majumdar}
\author[1]{Marc Stettler\corref{mycorrespondingauthor}}

\cortext[mycorrespondingauthor]{Corresponding author}
\ead{m.stettler@imperial.ac.uk}

\begin{abstract}
Passenger rejection by taxi drivers impacts the travel behaviour in many cities and suburban areas, often leaving those potential customers in non-popular zones stranded without access to taxis. To overcome this problem, many practices have been implemented, such as penalties to drivers, bans, and new pricing strategies. This paper presents a double auction taxi fare scheme, which gives both passengers and taxi drivers to influence the price, coupled with a clustering method to discourage strategic service rejection in the case study of Bangkok Metropolitan Region, Thailand, which has detailed data availability and uneven taxi journey distributions. The double auction mechanism is tailored to 2019 taxi trips, service rejection complaints, and local travel behaviour to boost transportation equity. To benchmark the performance of the new double auction scheme, a bespoke agent-based model of the taxi service in Bangkok Metropolitan Region at different rejection rates of 0\%-20\% was created. On one hand, the current rejection behaviour was modelled, and on the other, the double auction pricing strategy was applied. The results indicate that the double auction strategy generates a spatially distributed accessibility and leads to a higher taxi assignment success rate by up to 30\%. The double auction scheme increases pickups from locations that are 20-40 km from central Bangkok by 10-15\%, despite being areas of low profit. Due to the changing taxi travel landscape and longer taxi journeys, the total air pollutant emissions from the taxis increase by 10\% while decreasing local emissions within central areas of Bangkok by upto 40\%. Using a \textbaht 5 average surcharge, the total revenue drops by 20\%. The results show that an equity-driven pricing strategy as an implementation of transport policy would be beneficial.
\end{abstract}


\end{frontmatter}

% \linenumbers

\section{Introduction} \label{Sect1:Intro}
\begin{figure}[!ht]
\includegraphics[width=\textwidth]{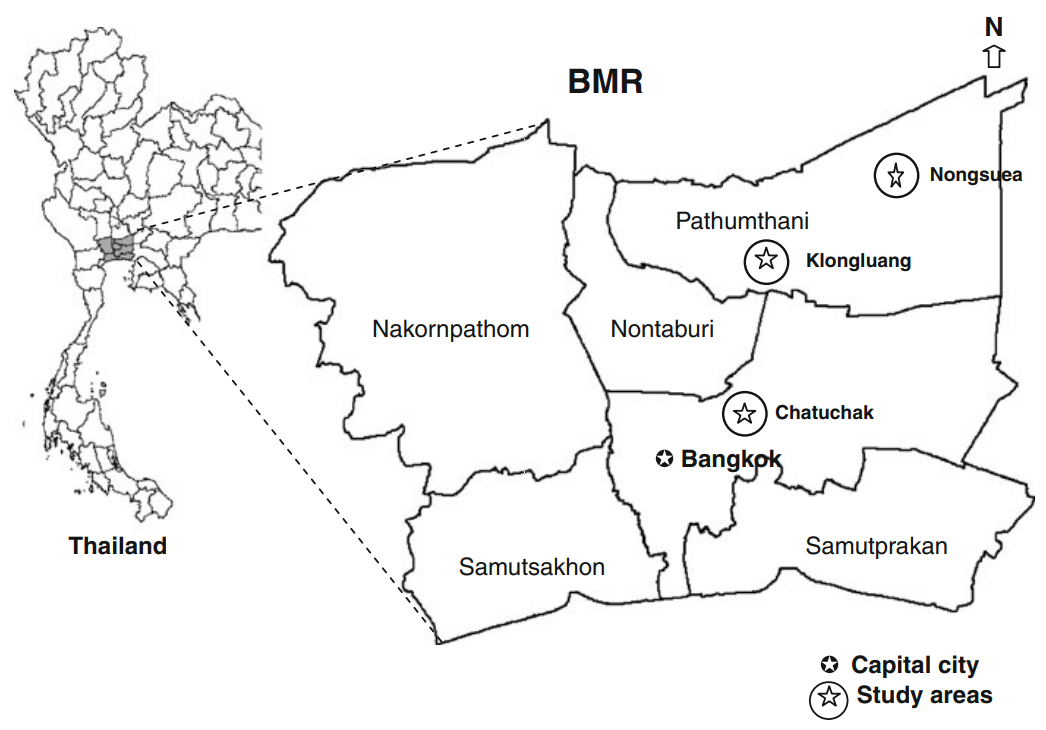}
\centering
\caption{Map of Bangkok Metropolitan Region}
\label{fig:map}
\end{figure}

Bangkok, Thailand, is a popular tourist destination and home to over 5.5 million residents in 2020 \cite{1_provinAdmin}. It sits within Bangkok Metropolitan Region (BMR) with five adjacent provinces of Nakhon Pathom, Pathum Thani, Nonthaburi, Samut Prakan, and Samut Sakhon \cite{2_atlas}, see Figure \ref{fig:map}.

\FloatBarrier

The BMR currently encompasses $7,762km^2$ after the city of Bangkok (referred to as Bangkok) expanded in all directions and many commuters travel from the neighboring provinces to work in the city of Bangkok leading to development of economic clusters and high-tech industrial and research areas \cite{3_IUDA}.

One of the main transport modes in the BMR is by taxi with a modal share of 18.9\% in 2015 \cite{4_Suparee}. The taxis are metered and currently use a fixed distance-based taxi fare scheme with various kilometre prices and a time-dependent component for standing and waiting times \cite{5_RegistrationTaxi}. Based on this fare system, trips below 20km are the most profitable for the taxi drivers, followed by medium-distance trips of 20 to 40 km with trips greater than 40km the least profitable\cite{6_Phiboonbanakit}. 

This has led to selective service provision by the taxi driver and leading to passenger rejection, or a refusal to carry a passenger based on their desired destination \cite{7_Nimmanunta}. In 2019, there were 21,124 complaints regarding passenger rejection within the BMR \cite{8_DeptLand}. The main rejected trips were to congested or less populated regions along with short rides under 50 baht, or \textbaht50 \cite{9_Atsawatheerasathien}. The local currency is Thai Baht, and \textbaht1 is equivalent to US \$0.03 on 29th of July, 2021 \cite{10_XE}. Atsawatheerasathien \& Kanitpong \cite{9_Atsawatheerasathien} characterized passenger-rejection behaviour in BMR through 952 passenger interviews during peak hours concluding that trips from Central to Eastern, from Northern to Southern, and from Northwestern to North Bangkok experienced higher rejection rates.

Moreover, the areas with a low-supply of low taxi service supply coincide with the regions of low-supply public transport areas, despite half of the BMR populations residing in this area \cite{11_Peungnumsai}. Peungnumsai et al. \cite{11_Peungnumsai} have also highlighted that regions with a high-income population had better public transportation access coverage in contrast to those with lower incomes.

Passenger rejection is not unique to Bangkok. In addition, there is an increasing use of taxi calling apps on smartphones globally. This provides new opportunities to change the taxi service landscape, particularly through new flexible, dynamic, real-time, and custom pricing strategies to be more inclusive to those living in traditionally less profitable areas. For example, in Prague, Czech Republic, the start-up Liftago and the Chinese company Didi have implemented a one-sided auction for providers or users to make bids, depending on the demand and supply levels. In Liftago, drivers bid to serve customer requests\cite{12_Liftago}, while in Didi, customers bid to get it served \cite{13_Loke}.

However, because Bangkok has areas of low demand and high supply and vice-versa, rather than a simple one-sided auction, double auction pricing mechanisms formed the baseline for development of a new pricing strategy to give both parties the autonomy to influence the price by disclosing what they are prepared to pay and therefore, to have a more efficient allocation of services.

This paper presents a newly developed taxi fare pricing scheme using a double auction scheme for Bangkok, Thailand using 2 year of taxi trip data and service availability. The paper assesses the performance of the new taxi fare pricing scheme for increased transportation equity and its impact on air quality by comparing to the current rejection behaviour using a bespoke agent-based model (ABM) of taxis and customers in the BMR. Rejection rates of 0, 5, 10, 15, and 20\% were modelled due to the uncertainty of the true rejection rate in Bangkok, as not all rejected trips may be reported.

The contributions of this work are summarized as follows:
\begin{enumerate}
\item A new combined double auction and clustering mechanism for app-based taxi fare pricing for better transportation equity has been designed,
\item a new ABM of BMR and the taxi services and customers has been built that can be adopted for any cities, and
\item the impact of transportation equity on total revenue and taxi vehicle emissions have been evaluated.
\end{enumerate}

The paper is organised as follows: A literature review is provided that covers previous work on double auction mechanisms, followed by a discussion of taxi demand estimation from various regions of BMR based on available taxi activity datasets. The following section presents the development of the double auction scheme for BMR and the corresponding rejection scheme based on current taxi landscape. This is followed by the description of the agent-based model. After that, the results are presented with discussions of comparing the two scenarios to show the benefits and downfalls of the double auction pricing scheme. The paper ends with suggestions for future work and conclusion.

% TODO: Find % of cars being Taxis in BMR to get the impact on emissions

\section{Background} \label{Sect2:Background}
While there are studies and implementations of use of road pricing and congestion charges to reduce emissions in Central cities overall, there is little study if at all done to changing pricing strategies to change the travel behaviour to reduce emissions in central areas of the cities.

% (TfL and Baghestani, New York, Chicago, etc.)

% Current pricing scheme in Bangkok taxi

\subsection{Double Auction}\label{Sect2:DoubleAuction}
A double auction is a decision process of sale and payment of goods or services, where both buyers and sellers make bids to either negotiate a compromise of a price, or otherwise the exchange does not take place \cite{14_Karamanis}. Although this mechanism can be seen in the stock exchange in practice, it is currently only theorised for ride-sourcing transportation, such as for ride-sharing operations \cite{14_Karamanis}, to the best of the authors’ knowledge. However, it has the potential to be widely adopted with the use of cloud-based apps for transport requests, faster and easily available mobile internet, and the increasing gig economy.

\subsubsection{Classical Double Auction mechanisms}
In a classical double auction process, there are multiple potential sellers and buyers of the taxi service. The potential buyers submit a bid of $ B = \{ b_1,{...},b_n \} $, and the potential sellers submit their offer of $ S = \{ s_1,{...},s_n \} $. The market then settles on the equilibrium price p, and buyers with bids higher than p and sellers who bid lower than p are considered for the win. Ultimately, based on a natural ordering of descending for buyers and ascending for sellers, the winners are the kth buyers, who offered the lowest bid above p, and the kth sellers, who offered the highest bid below p. The trading price p is determined by a mechanism that ideally satisfy the following four properties \cite{15_Kumar}:
\begin{enumerate}[1)]
\item Economic efficiency (EE) or allocative efficiency: The allocative efficiency is the ratio of the actual total profit of the auction participants and the total profit that can be obtained theoretically. This means that in a 100\% efficient double auction mechanism, the sum of all the prices of all players should be the highest possible. This is achieved when the bidding ends and the goods or service should be with the buyer who bid the highest and closest to the trading price, p, and this allows for the seller to receive highest possible income.
\item Individual rationality (IR): Every participant is rational and would not be participating in the auction if they could make a loss. The ideal property is that no participating agent can achieve negative utility in the auction, so their bids should follow the rule of $ S \leq  p \leq B $. 
\item Budget balance (BB): The budget in this case refers to the monetary transfers between the buyer and the seller. The more the monetary transaction is just between the buyer and seller with little or no financial deduction by the auctioneer or the taxi provider, the stronger the balanced budget. 
\item Truthfulness (TF), or incentive compatibility: Players bid for true values without the notion of other players’ bids \cite{15_Kumar}. 
\end{enumerate}  

These four properties, however, cannot be achieved simultaneously according to Myerson–Satterthwaite theorem \cite{16_Myerson}. There are some well-known mechanisms that meet some of the properties and already have been applied in other industries. Table \ref{tab:comparison} summarises the main findings.

% \topskip0pt
% \vspace*{\fill}
% text
% \vspace*{\fill}

% Add table here
\begin{table}[!ht]
\centering
\caption{Comparison of common double auction mechanisms}
\label{tab:comparison}
\begin{adjustbox}{width=1.68\textwidth, angle=90}
\begin{tabular}{|l|l|l|l|l|l|}
\hline
\multicolumn{1}{|c|}{Mechanism} &
  \multicolumn{1}{c|}{\begin{tabular}[c]{@{}c@{}}Price that the \\ winning buyer \\ pays, b\end{tabular}} &
  \multicolumn{1}{c|}{\begin{tabular}[c]{@{}c@{}}Price that the \\ winning seller \\ receives, s\end{tabular}} &
  \multicolumn{1}{c|}{\begin{tabular}[c]{@{}c@{}}Auctioneer receives \\ (negative indicates \\ auctioneer pays)\end{tabular}} &
  \multicolumn{1}{c|}{Advantages} &
  \multicolumn{1}{c|}{Disadvantages} \\ \hline
\begin{tabular}[c]{@{}l@{}}Vickrey–Clarke–Groves (VCG) \\ \cite{17_Georgiadis, 18_Babaioff} \end{tabular} &
    Buyer k pays $max(s_{k}, b_{(k-1)})$ &
  \begin{tabular}[c]{@{}l@{}}Seller k receives \\ $min(s_{(k-1)}, b_{k})$ \end{tabular} &
  $-(s-b)$ &
  \begin{tabular}[c]{@{}l@{}}The buyer, who values \\ the product the closest \\ to the trading price wins.\end{tabular} &
  \begin{tabular}[c]{@{}l@{}}There is no budget balance, \\ and the provider has \\ to subsidise the trade.\end{tabular} \\ \hline
Trade Reduction \cite{18_Babaioff} &
  Buyer k-1 pays $b_{k}$ &
  Seller k-1 receives $s_{k}$ &
  $b-s$ &
  \begin{tabular}[c]{@{}l@{}}The seller receives more \\ than their asking price.\end{tabular} &
  \begin{tabular}[c]{@{}l@{}}Weak budget balance with \\ the auctioneer has a surplus \\ and the mechanism is not \\ economically efficient.\end{tabular} \\ \hline
\begin{tabular}[c]{@{}l@{}}Dominant strategy \\ (also known as McAfee and \\ also used by Egan et al) \cite{19_McAfee,20_Egan_book}\end{tabular} &
  \begin{tabular}[c]{@{}l@{}}Buyer k pays $p=(b_{(k+1)}+s_{(k+1)})$ \\ if $s_{k}$\textless{}p\textless{}$b_{k}$\end{tabular} &
  \begin{tabular}[c]{@{}l@{}}Seller k receives $p=(b_{(k+1)}+s_{k+1)})$ \\ if $s_{k}$\textless{}p\textless{}$b_{k}$\end{tabular} &
  None if $s_{k}$\textless{}p\textless{}b\_k &
  \begin{tabular}[c]{@{}l@{}}This is an extension of \\ the trade reduction \\ mechanism.\end{tabular} &
  \begin{tabular}[c]{@{}l@{}}If the buyer and seller k \\ don’t bid around p, \\ the budget balance is weak, \\ and the mechanism is not \\ economically efficient.\end{tabular} \\ \hline
\begin{tabular}[c]{@{}l@{}}Efficient Double Auction \\ (developed by Zhou and Xu) \cite{21_Zhou}\end{tabular} &
  \begin{tabular}[c]{@{}l@{}}Introduce fictitious buyer k+1, so \\ Buyer k pays 
  $p=average($ \\
  $min(b_{k},s_{k+1}), max(s_{k},b_{k+1}))$ \end{tabular} &
  \begin{tabular}[c]{@{}l@{}}Introduce fictitious seller k+1, so \\ Seller k receives \\
  $p=average(min(b_{k},s_{k+1}), max(s_{k},b_{k+1}))$\end{tabular} &
  None &
  \begin{tabular}[c]{@{}l@{}}This is an adaptation of \\ the dominant strategy\end{tabular} &
  \begin{tabular}[c]{@{}l@{}}This strategy assumes there \\ is a preference on winning, \\ so a trade is more likely to \\ happen at a cost of lack \\ of truthful bids.\end{tabular} \\ \hline
\end{tabular}
\end{adjustbox}
\end{table}
\FloatBarrier

\subsubsection{Double Auction mechanisms in the taxi system}

In the context of taxi systems, the goods are taxi services, and the trading prices are not absolute values but the unit price of a trip. Egan et al. \cite{22_Egan} built a mechanism based on the McAfee’s mechanism, in which agents bid for the price of the entire trip through McAfee’s Double Auction using markets with a similar trip and pickup distances. Zhou and Xu \cite{21_Zhou} pointed out that the Egan et al. \cite{22_Egan} approach affects market flexibility precluding potential deals that do not match theoretical scenarios. Consequently, they opted to base the auction on the unit price of a trip. Zhang, Wen and Zeng \cite{23_Zhang} designed a discounted trade reduction mechanism in which bids are adjusted, hereby named discounted, with tunable factors accounting for the customers' waiting time and of the agent’s reputation. 
By introducing fictitious bidders so that demand and supply curves are forced to cross, the Efficient Double Auction mechanism from Zhou and Xu considers all potential trades and no surplus by the marketplace from trading is generated \cite{21_Zhou}. Agents in this mechanism were proven to take individually rational decisions and follow a truthful strategy for all passengers and drivers with winning preferences with non-negative utility. In addition, agents in an urgent situation can overbid to ensure winning the auction, while risking lowering their utility. 
In the literature reviewed, due to the ability to include all potential bidders, the resilience to overbidding, and meeting the properties of individual rationality, budget balance and economic efficiency has made the Efficient Double Auction scheme from Zhou and Xu \cite{21_Zhou} the best fit for BMR. However, one of the difficulties of application of the double auction in taxi rides is the inclusion of long-term strategies for providers, which contradicts the short-term gains of the budget balance property of double auction. In addition, there is a reliance of provision of unrestricted financial resources by taxi users, which will in turn favor the buyers in high-income areas, where there already is an abundance of taxi supply \cite{22_Egan}. To overcome this, a new clustering method is proposed.
In terms of double auction in taxi systems, there are two classes: static and online \cite{22_Egan}. The former uses information of all the passengers and drivers within the system; whereas the latter allows for passengers and drivers to arrive at different times. In developing our own double auction mechanism, the requirement of a new clustering method and the online case have been considered.

\section{Materials and Methods}
\subsection{Road Network Data}

The road network information of the region of BMR was obtained from OpenStreetMap \cite{24_OpenStreetMap}. Only the main roads and highways were adopted for the simulation to simplify the network.

\subsection{BMR Taxi Data}
Taxi GPS (Global Positioning System) tracking system data were provided by the Intelligent Traffic Information Center Foundation (iTIC), Thailand \cite{25_iTICFoundation}. From this data, it was estimated that approximately 5,300 taxis were operating per day between January and April 2019. The frequency of GPS signal transmission varies between 1 to 3 minutes, depending on the status of taxis internal combustion engines. Each GPS data consists of the following: vehicle identification, latitude, longitude, speed, direction, occupied or unoccupied, and active or inactive engine status. After generating a Voronoi Polygon mesh from the BMR network nodes and matching the corresponding GPS points, an average hourly link speed was estimated from the speeds associated with the initial and final link nodes. The accepted trips origin/destination (O/D) matrix was obtained from Bangkok taxi probe data \cite{25_iTICFoundation}. Overall, demand is higher between 6 am and 11 pm of the order of 8,000 accepted trips per hour in the BMR. 

\subsection{Trip Rejection Rate} \label{rejection_rate}
By combining the demographic statistics in 2019 derived from the Bureau of Registration Administration \cite{1_provinAdmin}, network speed obtained from the estimation of hourly average speed using taxi GPS data \cite{25_iTICFoundation}, and public transport accessibility \cite{11_Peungnumsai}, the rejected demand was estimated to add to the base demand captured by the taxi GPS data \cite{25_iTICFoundation, 26_Siangsuebchart}. This sum forms the potential demand, which is the input for the ABM. The average rejection rate is the probability a trip is rejected, considered homogeneous throughout the network. The uncertainty around that parameter fostered the evaluation of five potential demand levels based on the following rejection rates: 0, 5, 10, 15 and 20\%. That range was chosen after comparing base-demand total requests at around 10,000 requests per hour in the entire region to the Transport Statistics Report for 2019 and 2020 \cite{8_DeptLand}, which accounts for the officially reported complaints of the taxi service.
% \todo{won't you say something about the overal structure of the methodology. wouldn't it be nice to have a flowchart/diagram of this work? ie. describe the overal way that your methodology works, with the components and the overall assumptions? and then you go through the description of the various units}

\subsection{Calculation of rejected trips}
The methodology of calculating the trip rejection data is structured into two parts: 1) overall average trip rejection, 2) district-specific trip rejection. For the first part, the total accepted trips, as shown in BMR Taxi data, have been multiplied by the average rejection rate $\beta = [0\%, 5\%, 10\%, 15\%, 20\%]$ from \ref{rejection_rate} to develop the intermediate O/D matrix. Secondly, the O/D matrix was further multiplied by the likely regional rejection factor, which were formed of public transport access disparity and population density of various districts. 

Congestion was characterized by network speed due to the lack of road infrastructure data. For each district $x$ and hour $t$, the mean hour speed was estimated and then divided by 120 km/h, Thailand's official speed limit, to obtain a speed factor, $\bar{v}(x,t)$ between 0 and 1, where 0 represents congested regions. A public transport accessibility factor, $\bar{g}(x)$) between 0 and 1, 1 meaning a high gap, was deduced from \cite{11_Peungnumsai}. Speed and public transport access factors generated a network factor, $F(x,t)$ through Equation 1:

\begin{equation} \label{eq:1}
F(x,t) = exp[-\eta_v \cdot \bar{v}(x,t)] \cdot exp[\eta_g \cdot \bar{g}(x)]
\end{equation}

where $\eta_v$ = weight factor, 0.9, and $\eta_g$ = public transport accessibility weight factor, 1.2, which is greater than $\eta_v$ to emphasize the lack of transport equity.

The district rejection factor $\delta(x,t)$ is obtained as shown in Equation \ref{eq:2}:

\begin{equation} \label{eq:2}
\delta(x,t) = \frac{F(x,t)}{pop(x)}
\end{equation}

where $pop(x)$ is the ratio of population in each district $x$ over the total population.

The rejected trips O/D matrix and trip rejection probability is obtained from Module 3. A trip rejection factor $\chi$ between two districts $x_{ini}$ and $x_{end}$ at a time $t$ is as shown in Equation \ref{eq:3}:

% change ini and end to i and j
\begin{equation} \label{eq:3}
\chi(x_{ini},x_{end},t_{ini}) = \delta(x_{ini},t) \cdot \delta(x_{end},t)
\end{equation}

where $\delta(x_{ini},t)$ = origin district rejection factor and $\delta(x_{end},t)$ = destination district rejection factor.

By fitting $\chi$ to a [1,2] range, the adjusted trip rejection factor, $\bar{\chi}(x_{ini},x_{end},t_{ini})$ is obtained. Finally, the rejection probability of a trip between districts $x_{ini}$ and $x_{end}$ at a time, $t$, $Pr[Rejection|x_{ini},x_{end},t]$, is as shown in Equation \ref{eq:4}.

\begin{equation} \label{eq:4}
Pr[Rejection|x_{ini},x_{end},t] = \beta \cdot \bar{\chi}(x_{ini},x_{end},t_{ini})
\end{equation}

The rejected trips O/D matrix, $T(x_{ini},x_{end},t_{ini})$, is the product of the total rejected trips obtained in Module 1 times the share of rejected trips associated with each O/D pair, noted as $\Theta(x_{ini},x_{end},t_{ini})$, as shown in Equation \ref{eq:5}.

\begin{equation} \label{eq:5}
T(x_{ini},x_{end},t_{ini}) = round[\Theta(x_{ini},x_{end},t_{ini}) \cdot {Total \ rejected \ trips}]
\end{equation}

where $\Theta(x_{ini},x_{end},t_{ini})$ = share of rejected trips associated with each O/D pair, calculated as shown in Equation \ref{eq:6}:

\begin{equation} \label{eq:6}
\Theta(x_{ini},x_{end},t_{ini}) = \frac{\zeta(x_{ini},x_{end},t_{ini})}{\sum{\zeta(t_{ini})}}
\end{equation}

where 

\begin{equation} \label{eq:7}
\zeta = 
\begin{cases}
    \bar{\chi}(x_{ini},x_{end},t_{ini}) & \text{if} \ \bar{\chi}(x_{ini},x_{end},t_{ini}) > Q_{3} \\
    0 & \text{otherwise}
\end{cases}
\end{equation}

where $\bar{\chi}(x_{ini},x_{end},t_{ini})$ = adjusted trip rejection factor and $Q_{3}$ = 3rd quartile of $\bar{\chi}$.

%%% Look for Figure 2

\subsection{Development of Pricing Schemes}

\subsubsection{Double Auction for the BMR}

The newly developed double auction algorithm emulates an app-based double auction matching mechanism for taxi services. Customers and drivers bid for a trip surcharge in Thai Baht, \textbaht, to be added on top of the current fare scheme, as modelled by Pueboobpaphan, Indra-Payoong \& Opasanon to indicate trip rejection \cite{27_Pueboobpaphan}. 

Initially, if the number of customers exceeds the number of available vehicles by a factor $\chi_{c/v}$ =  2, the same as the rejection model, only the oldest $\chi_{c/v}\cdot V$ requests will enter double auction algorithm, where $V$ is the total number of available vehicles at a given timestep.

\begin{figure}[!ht]
\includegraphics[width=\textwidth]{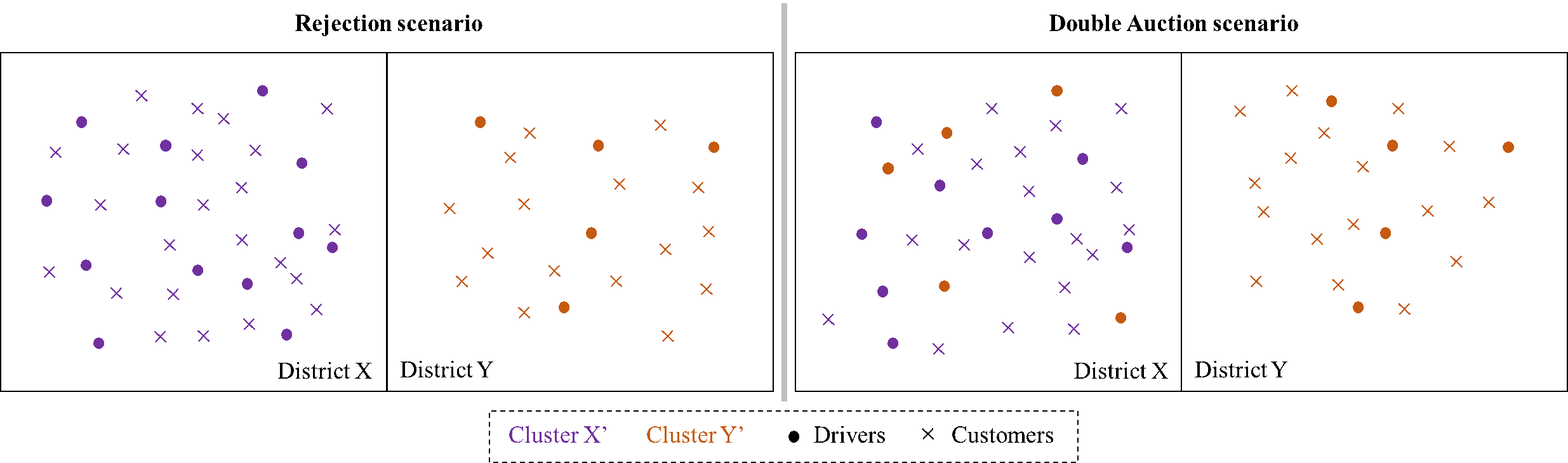}
\centering
\caption{New clustering method for double auction algorithm}
\label{fig:clustering}
\end{figure}

With the double auction algorithm, a new clustering method is introduced. The rejection scenario, described in the next section, groups agents with similar origins and destinations by proximity based on the k-means algorithm to minimise pickup distance, which may exclude potential matchings of agents in areas with concentrations of customers or drivers, as shown in Figure \ref{fig:clustering}. The newly introduced clustering method with the double auction algorithm introduces taxis to seek customers in nearby clusters as its own and allocates taxis to critical clusters with a much higher demand, as shown in Figure \ref{fig:clustering}. This allows for the double auction method to pursue a balanced submarket to homogenize assignment probabilities. 

\FloatBarrier

An average cluster size $\gamma$, is fixed at 20 customers following Zhou and Xu \cite{21_Zhou}, and a maximum number of clusters $\bar{C}$ is set to 50. Clusters with less than 20 requests are discarded. Yet, customers in the discarded clusters with waiting times greater than 20 minutes or having played more than 7 auctions are reallocated to the closest non-discarded cluster, which will undergo an auction. The remaining requests are left unassigned. 
To enforce balanced submarkets with similar numbers of customers and drivers, the maximum number of drivers in a cluster is limited to $V_{lim}$, as shown in Equation \ref{eq:8}. After sorting vehicles by their last customer drop off time, they are allocated to the closest available cluster. 

\begin{equation} \label{eq:8}
V_{lim} = N \cdot \eta_{v/c}
\end{equation}

where $N$ = number of customers in a cluster and $\eta_{v/c}$ = ratio of vehicles per customer, set to 1.1.

This algorithm ensures that all vehicles have the chance to bid for a customer, that all clusters have balanced numbers of bidders to homogenize service accessibility and that vehicles with longer waiting times are allocated to the closest clusters. 

The formed clusters pursue a double auction mechanism following Zhou and Xu \cite{21_Zhou} algorithm. Bids are simulated from a normal distribution $N(\mu,\sigma)$. For both customers and drivers, $\sigma$ = 0.3. An average bid $\widetilde{\mu}$ is defined for all agent. Following an initial proposal by the Government of Thailand \cite{8_DeptLand}, $\widetilde{\mu}$ was fixed at \textbaht5, despite only representing a 2.5\% rise in a common 30km trip costing \textbaht200.

Each customer has a different average bid $\mu_{c}$, which is calculated using Equation \ref{eq:9}, dependent on a bid factor $\eta_{t}$, calculated as shown in Equation \ref{eq:10}, accounting for the trip’s rejection probability:

\begin{equation} \label{eq:9}
\mu_{c} = \eta_{t} \cdot \widetilde{\mu}
\end{equation}

\begin{equation} \label{eq:10}
\eta_{t} = 1 + Pr[Rejection]
\end{equation}

The average drivers' bid $\mu_d$, as calculated in Equation \ref{eq:11}, is constant for drivers in the same cluster. Their profit-oriented view turned $\mu_d \geq \bar{\mu}_{c}$, being $\bar{\mu}_c$ the cluster average customers bid.

\begin{equation} \label{eq:11}
\mu_{d} = \eta_{d} \cdot \bar{\mu}_{c}
\end{equation}

where $\eta_{d}$ = ratio between customers and drivers' mean bid, 1.1.

Bids were bounded to ensure they fall within a reasonable range imposing $\eta_{max}$ = 1.75 and $\eta_{min}$ = 0.5, as shown in Equation \ref{eq:12}.

\begin{equation} \label{eq:12}
-\eta_{min} \cdot min(\mu_{c}, \mu_{d}) \geq bid \geq \eta_{max} \cdot max(\mu_{c},\mu_{d})
\end{equation}

\begin{figure}[!ht]
\includegraphics[width=\textwidth]{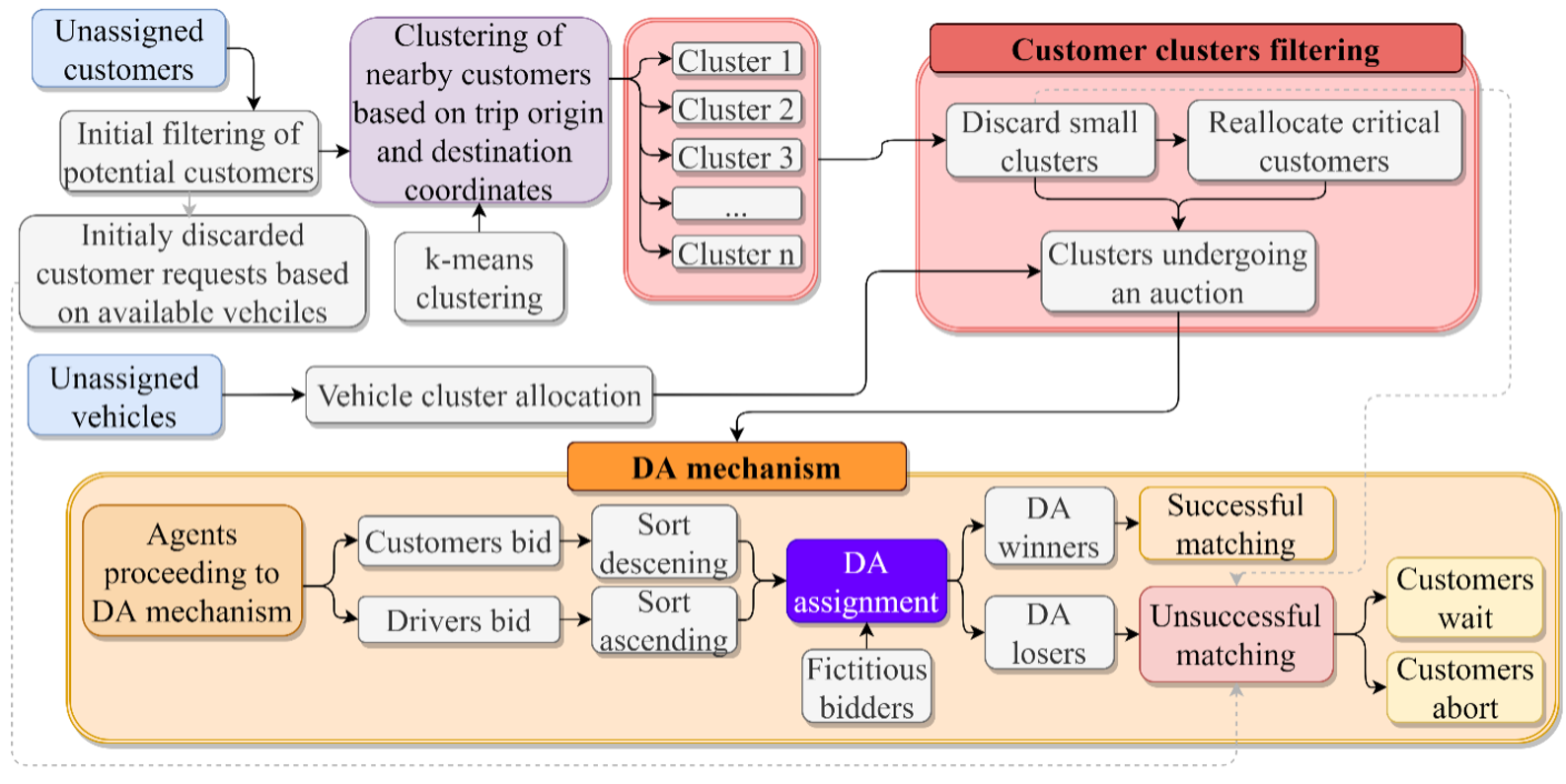}
\centering
\caption{Double auction algorithm diagram}
\label{fig:doubleAuction}
\end{figure}

The outputs are assigned and unassigned vehicles and customers. Request abortion happens for unassigned customers with waiting times above 20 minutes or that have played more than seven auctions \cite{7_Nimmanunta}. A summary of the double auction algorithm combined with the new clustering method is shown in Figure \ref{fig:doubleAuction}.

\FloatBarrier

\subsubsection{Rejection Model}
This algorithm emulates an app-based taxi service in which drivers strategically reject certain request and the current fare scheme model,as developed by Pueboobpaphan, Indra-Payoong \& Opasanon to model rejection \cite{27_Pueboobpaphan}. Initially, through a k-means algorithm, clusters of customers and drivers are generated based on their origin location. The number of clusters $C$ is obtained as shown in Equation \ref{eq:13} and is limited to $\bar{C}$=50. 

\begin{equation} \label{eq:13}
C = \frac{Total \ customers \ and \ vehicles}{\gamma} < \bar{C}
\end{equation}

where $\gamma$ = average cluster size of 30 agents.

Customers and vehicles in each cluster are sorted based on their waiting time. The first $n$ customers and vehicles are provisionally assigned, with $n$ being the minimum between the total number of customers and of vehicles. Then, drivers can either accept or reject the request, as per Equation \ref{eq:14}, according to the associated trip rejection probability obtained in section \ref{rejection_rate}.

\begin{equation} \label{eq:14}
{match} = 
\begin{cases}
    reject & \text{if} \ \xi < Pr[Rejection|x_{ini},x_{end},t] \\
    accept & \text{if} \ \xi \geq Pr[Rejection|x_{ini},x_{end},t]
\end{cases}
\end{equation}

where
$\xi$ = random number between 0 and 1.

\begin{figure}[!ht]
\includegraphics[width=\textwidth]{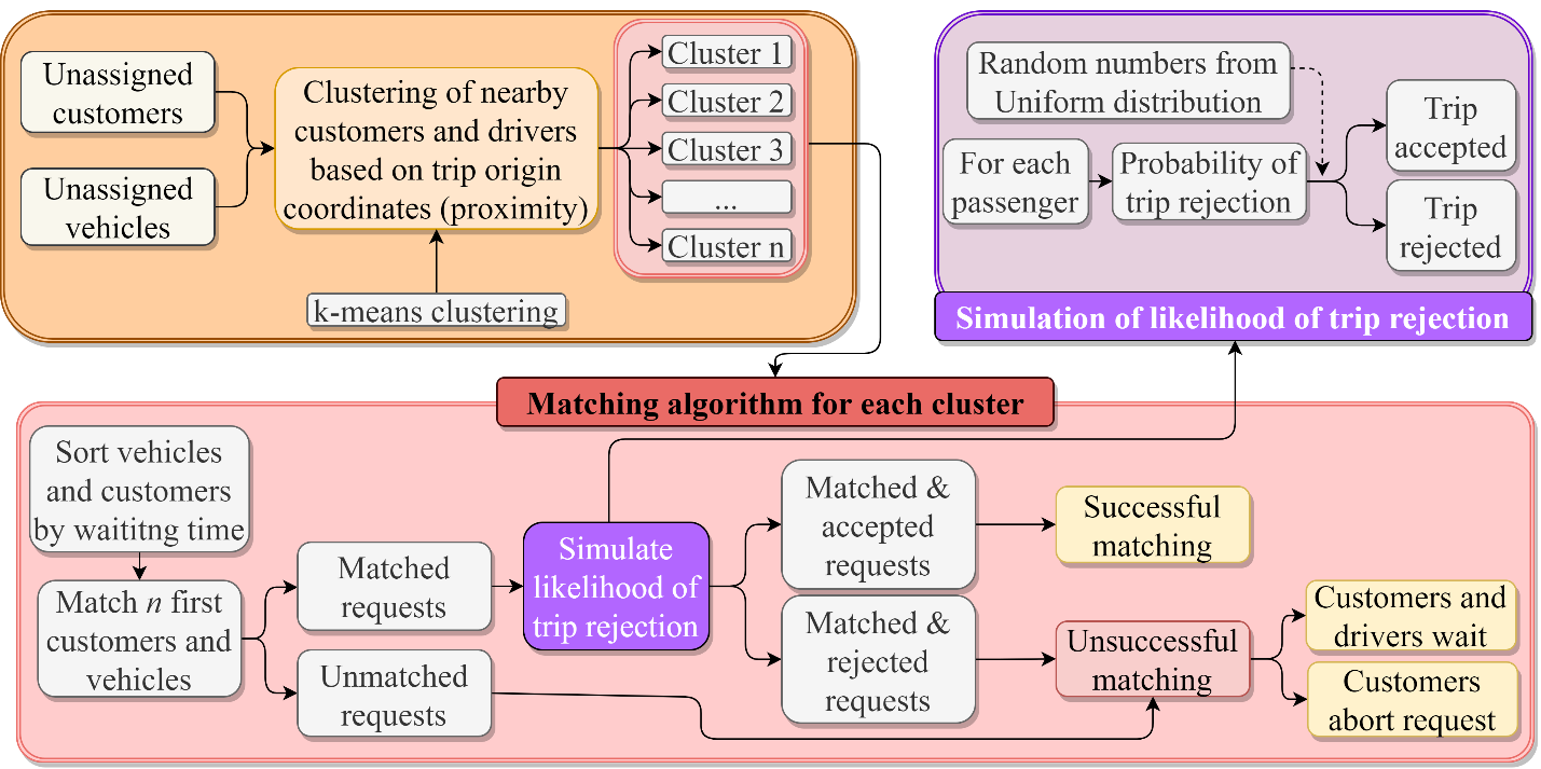}
\centering
\caption{Rejection algorithm architecture}
\label{fig:rejection}
\end{figure}

The outputs are assigned and unassigned vehicles and customers. Request abortion happens for unassigned customers with waiting times above 20 minutes or that have been rejected more than 7 times. A summary of the rejection algorithm is shown in Figure \ref{fig:rejection}.

\FloatBarrier

\subsubsection{Performance Indicators}
Two new performance indicators were used to compare the performances: balanced markets and service accessibility. Balanced markets refer to clusters with a similar number of customers and vehicles. Service accessibility is the ratio between the number of vehicles to which a customer can be potentially assigned over the number of customers in that cluster or submarket. Balanced submarkets homogenize service accessibility. 

The supply $S$ offered to a given district throughout the study period is the average number of available taxis in a district during a minute. Similarly, the demand level $D$ is the average total taxi requests in a district during a minute. Their difference is the gap in supply and demand $G$. 

The success rate for a taxi driver to be assigned a customer is the mean of the ratio of assigned taxis among the total unassigned ones in a timestep. Total revenues ($TR$), as shown in Equation \ref{eq:15}, are derived from an equivalent constant distance-based price rate $\bar{R}$ (in \textbaht/km) estimated from the average trip distance using the current taxi fare. Double auction also includes the revenues from the variable surcharge, which are the sum of all individual trading prices $V_{DA}$ of the assigned trips, from which the average trading price $\bar{V}_{DA}$ is obtained.

\begin{equation} \label{eq:15} 
TR = F \cdot \bar{D}_{1c} \cdot \bar{R} + \sum V+{DA}
\end{equation}

where $\bar{D}_{1c}$ = total average distance travelled with at least 1 customer per vehicle and $F$ = fleet size.

The average trip price $\overline{TP}$, as shown in Equation \ref{eq:16}, is:
\begin{equation} \label{eq:16} 
\overline{TP} = \bar{D}_{1c} \cdot \bar{R} + \bar{V}_{DA}
\end{equation}

\subsection{Agent-based Model}
\begin{figure}[!ht]
\includegraphics[width=0.5\textwidth]{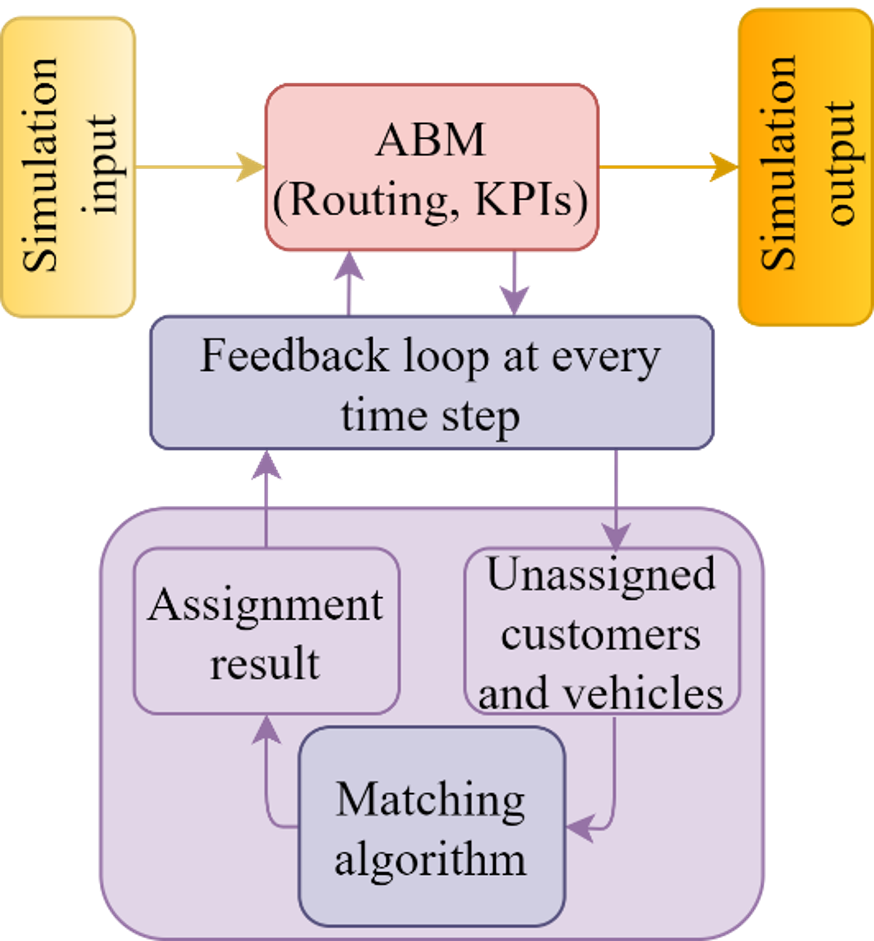}
\centering
\caption{ABM feedback loop}
\label{fig:abm}
\end{figure}

In order to assess the performance, a new agent-based model of BMR was created. Due to its modular design, just the matching algorithm could be interchanged. At each time step, the feedback loop that assigns vehicles with customers using either double auction or rejection algorithm was performed and updated the agents’ attributes. For each timestep, the loop inputs are unassigned customers and vehicles; the output is matched customers and vehicles, unassigned vehicles, unassigned customers that decide to wait for a new assignment and customers that abort the request and leave the simulation. To route the taxis the A* routing algorithm was applied. The ABM architecture is shown in Figure \ref{fig:abm}.

\FloatBarrier

The study period ranged from 6 am to 10 pm on a weekday with timesteps of one minute, which includes a warm-up and a cool-down period of 1 hour each. Vehicles in all the scenarios were randomly allocated using the same random seed. The demand level varied throughout the day based on the taxi data \cite{25_iTICFoundation} and calculation methodology as described in section BMR Taxi Data.

\subsection{Emissions models}

To understand the implication of the new pricing strategy on the environment, the emissions of the taxis have also been modelled. Approximately, 75\% of taxis in BMR are bi-fuel CNG-Gasolines and 25\% are bi-fuel LPG-Gasolines. Taxi vehicles in Thailand normally follow EURO 4 standards \cite{27_Pueboobpaphan} and a popular model for taxis is the Toyota Corolla Altis 2008-2013 1.8 G CNG \cite{28_Uttamang}.
Based upon the literature review \cite{26_Siangsuebchart, 27_Pueboobpaphan, 28_Uttamang, 29_CarDekho, 30_Cheewaphongphan, 31_smit, 32_Boulter, 33_Zhai} and n the model’s contemporaneity and range of applicability, COPERT \cite{34_Gkatzoflias} was chosen as the vehicle emissions model used in this analysis for the following reasons. Firstly, bi-fuel Gasoline-LPG and gasoline-CNG vehicles fit COPERT’s bi-fuel category. Secondly, Thailand standards to classify vehicles are based on European standards, inducing a major correspondence with COPERT regulatory classes. Thirdly, the outputs from the ABM are vehicle positions at different times, hence the speed is available, which matches the inputs required by COPERT.
The COPERT emission factors for CNG/LPG-gasoline bi-fuel, small segment, Euro 4 standard show that CO and Volatile Organic Compounds (VOC) increase with speed, NOx’s reduce, and PM emissions augment with speeds lower than 5km/h.

% %TODO
% \textcolor{red}{Manlika, can you check the percentage of taxis of overall traffic in BMR in 2019, please? I think the page \url{https://web.dlt.go.th/statistics/} supposedly is it but it's in Thai}

\section{Results} \label{Sect5:Results}
\begin{figure}[!ht]
\includegraphics[width=\textwidth]{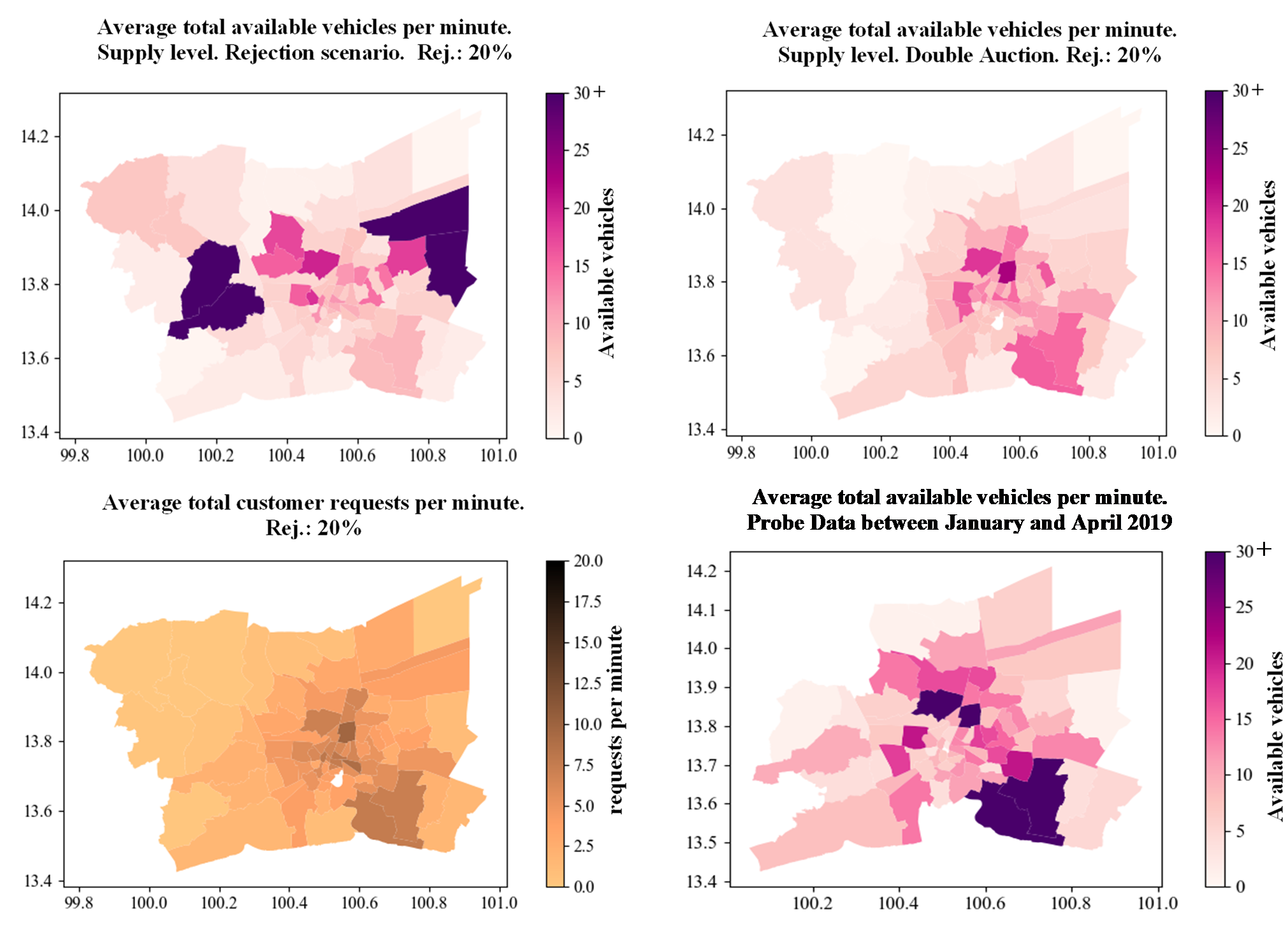}
\centering
\caption{Rejection scenario supply level (top left), double auction scenario supply level (top right), potential demand level (bottom left) for rejection rate of 20\%, and probe data supply level (bottom left)}
\label{fig:rejectionscenario}
\end{figure}

The results from the ABM simulations of both the rejection and double auction scenarios for rejection rate of 20\% have been portrayed in terms of the taxi supply in Figure \ref{fig:rejectionscenario}. These are co-portrayed with the potential demand level based on 20\% rejection rate calculated as described in sub-subsection \ref{rejection_rate}, and the actual supply based on the taxi GPS data \cite{25_iTICFoundation}.  

\FloatBarrier

When compared to the potential demand, the taxi supply following the double auction pricing strategy is proportional distributed by increasing services in the mid-distance high-demand yet low-supply areas by 50\%, while decreasing supply in other areas compared to the rejection scenario by 50\%. Compared to the actual taxi demand data, the rejection scenario shows similar areas of supply.

\begin{figure}[!ht]
\includegraphics[width=\textwidth]{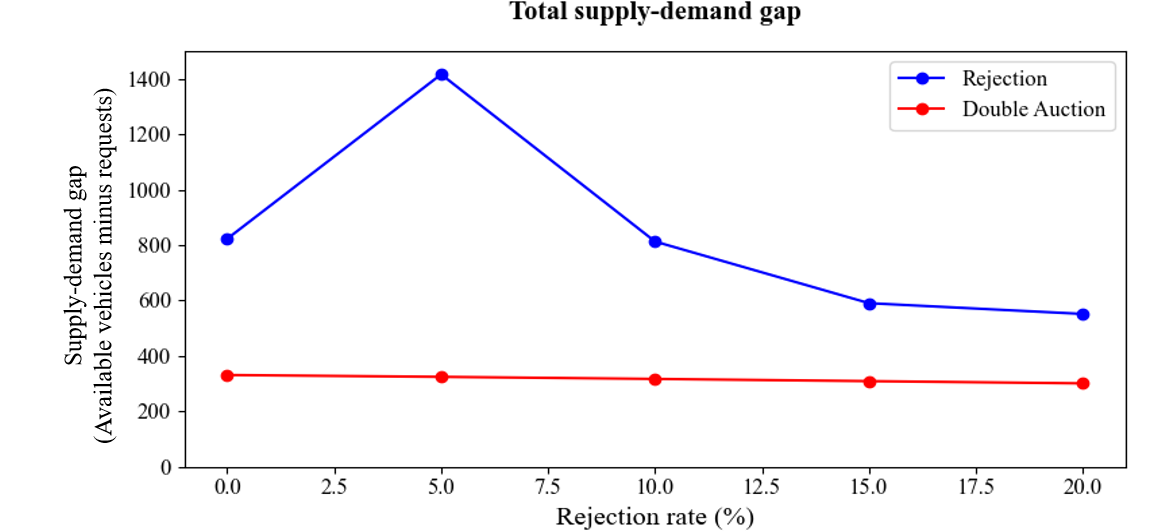}
\centering
\caption{Supply-demand gap based on the rejection and double auction scenarios}
\label{fig:supply_demand}
\end{figure}

The resulting the total demand-supply gap shows that through the double auction pricing strategy, this gap could be reduced, and the taxi service accessibility and transportation equity could be increased, as shown in Figure \ref{fig:supply_demand}. The 5\% rejection scenario experienced the maximum gap and narrowed with greater demand and rejection rates. 

\FloatBarrier

% FIGURE 8 Supply-demand gap based on the rejection and double auction scenarios

\begin{figure}[!ht]
\includegraphics[width=\textwidth]{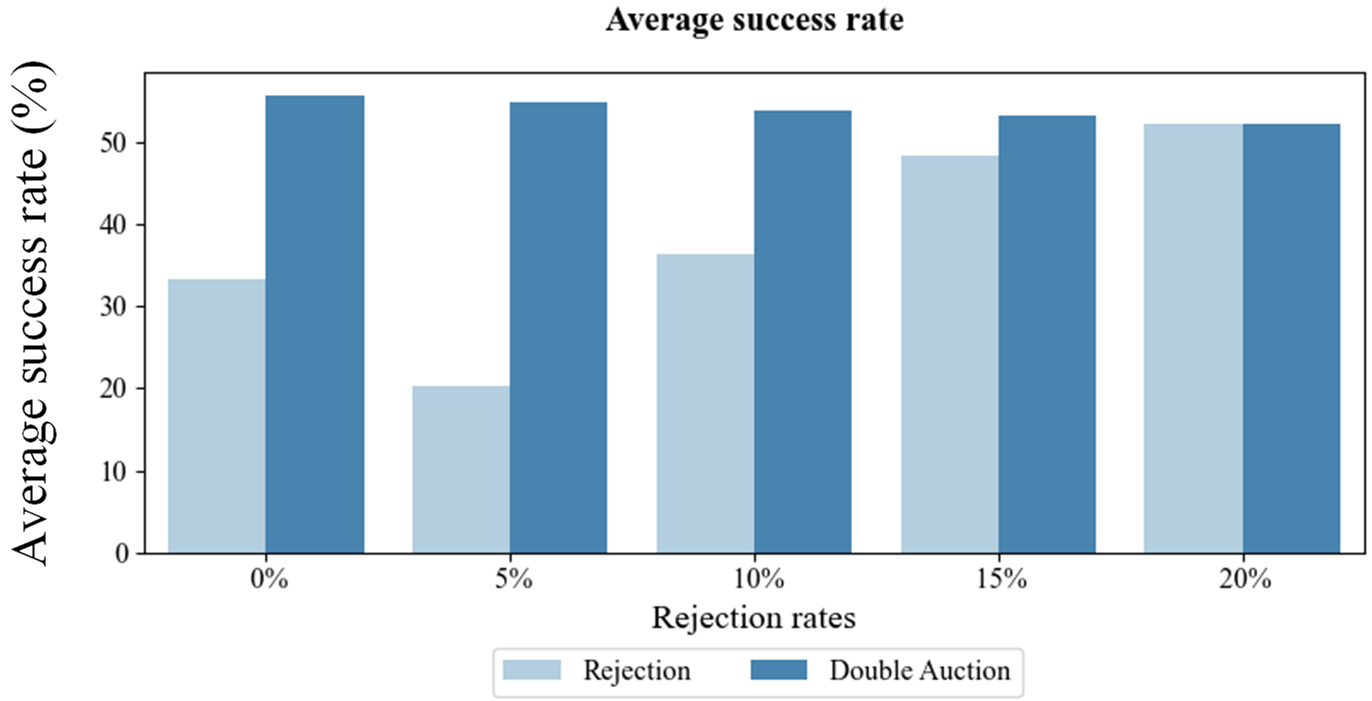}
\centering
\caption{Average matching success rate based on the rejection and double auction scenarios}
\label{fig:matching}
\end{figure}

Moreover, with the double auction strategy, the average success rate remained constant at 53\% for all rejection rates, as shown in Figure \ref{fig:matching}. In contrast, the average success rate was significantly lower for the rejection scenarios, except for the in the 20\% rejection rate. The minimum average success rate for the rejection scenario was at the 5\% rejection rate.

\FloatBarrier

% FIGURE 9 Average matching success rate based on the rejection and double auction scenarios

\begin{figure}[!ht]
\includegraphics[width=\textwidth]{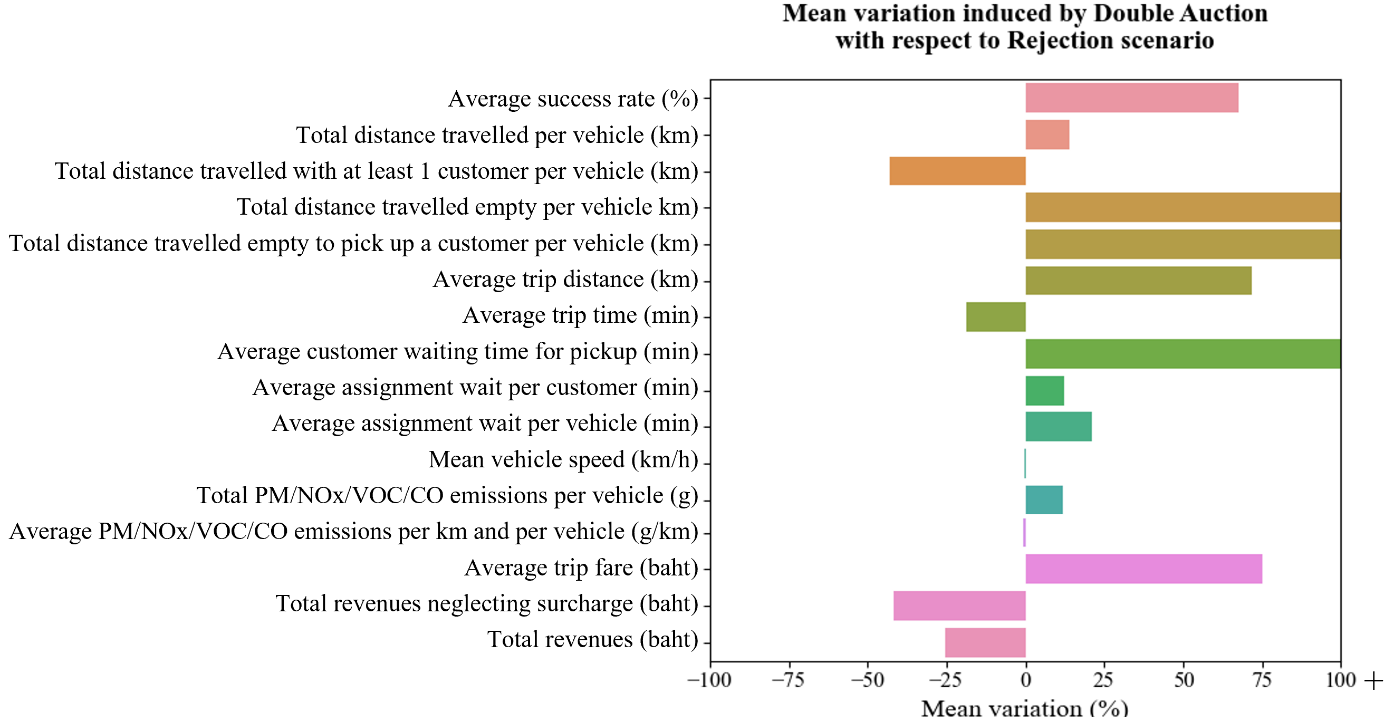}
\centering
\caption{Summary of the results and comparison between rejection and double auction scenarios}
\label{fig:results}
\end{figure}

The above translated into the double auction scenario observing 5 to 10\% additional pickups in mid-distance districts from Central Bangkok, where the origin was 20 to 40km away from the central areas of Bangkok city. Yet, the pickups decreased 15-20\% in the centre and over 25\% in remote areas. This led to average pickup distance doubling, increasing the total distance travelled empty by 180\%. The total travelled distance per vehicle supplemented 13\%, passing from 360km in rejection scenario to 400km. Additionally, as most trips assigned in Central Bangkok are shorter, double auction scenario captured longer trips of average of 60km versus 35km, yet with a shorter duration of average of 60 minutes versus 70 minutes. Consequently, 40\% fewer customers were served, passing from 70,000 trips in rejection scenario to 40,000. Both scenarios evoked similar average assignment waiting times of 10 minutes. Pickup time more than tripled, passing from 8 to 27 minutes due to the increase in pickup distances. All aborts in rejection scenario were due to excessive waiting time, whereas in the double auction scenario, the main cause was the excessive number of auctions played. The average trading price ranged \textbaht5 to \textbaht6 per trip progressively from 0 to 20\% scenarios. The higher trip distance raised the average trip fare, \textbaht350 against \textbaht200 in rejection model. Nonetheless, total revenues lowered 20\%, passing from \textbaht8 million to \textbaht6 million. A summary of the presented results is illustrated in Figure \ref{fig:results} 

\FloatBarrier

% FIGURE 10 Summary of the results and comparison between rejection and double auction scenarios

\subsection{Emissions Implications}
\begin{figure}[!ht]
\includegraphics[width=\textwidth]{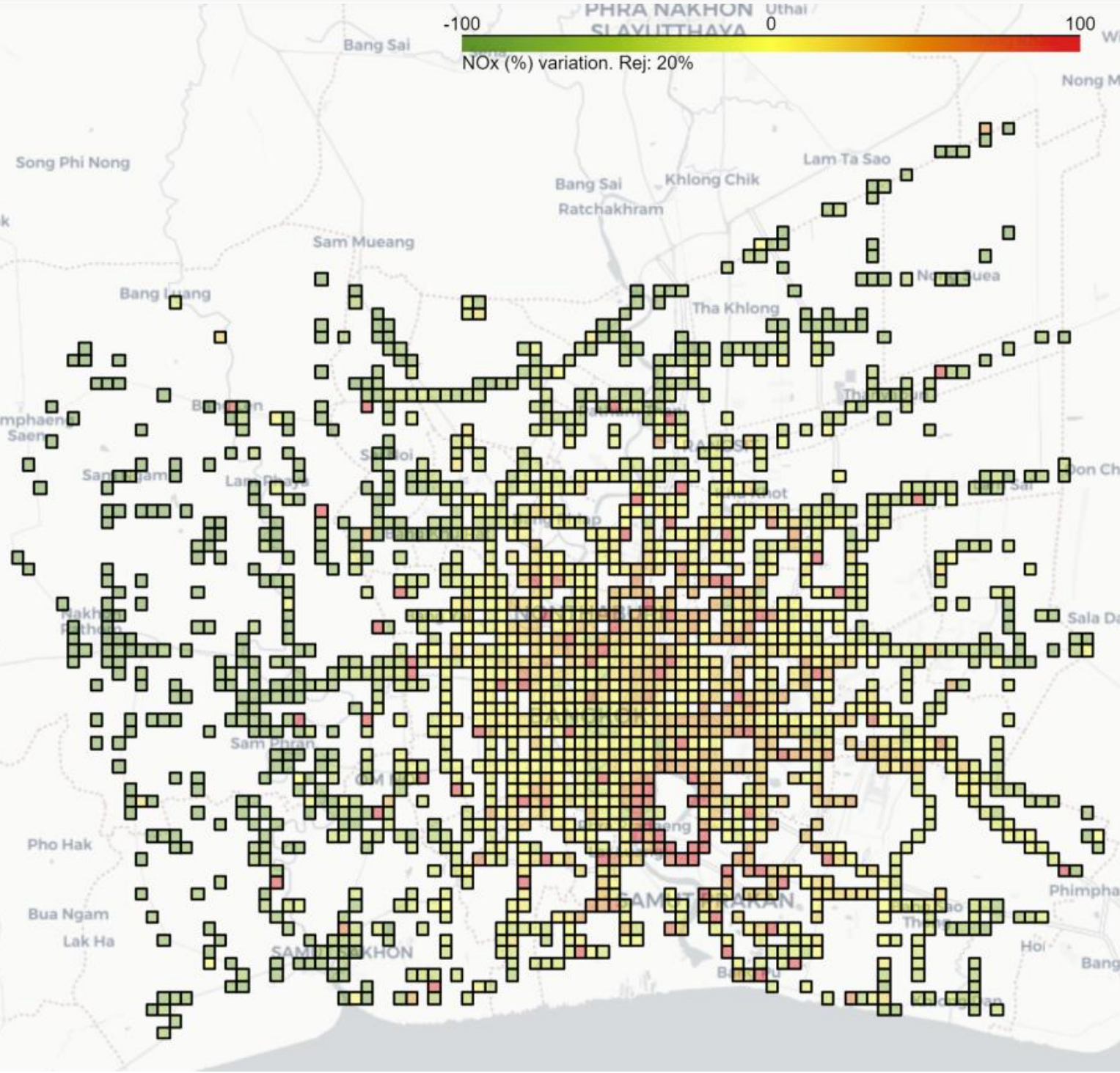}
\centering
\caption{Percentage variation in total NOx emissions considering 1km x 1km square grid. Rejection rate of 20\%.}
\label{fig:Nox}
\end{figure}

Between the rejection and double auction scenarios, the overall emissions increased by 10\%. Yet, dispersing the fleet onto the suburbs decreased 20\% to 40\% emissions in Central Bangkok linked to a 10\%-20\% growth in surrounding districts due to the higher supply. The example of differences of local NOx emissions between rejection and double auction scenario is shown in Figure \ref{fig:Nox} as a grid of 1km x 1km, where green shows decrease in emissions compared to the rejection model.

\FloatBarrier

% FIGURE 11 Percentage variation in total NOx emissions considering 1km x 1km square grid. Rejection rate of 20\%.

\section{Discussion} \label{Sect6:Discussion}
The trade-off between empty driving distance for picking up customers and service accessibility explains the capacity of the double auction to dynamically adapt to the real demand level of the region by mobilizing the taxi fleet proportionately to the number of requests and increasing accessibility. Favouring transportation equity with double auction comes at the cost of tripling pickup distance and time and 15-20\% fewer pickups in central Bangkok to better serve historically under-supplied districts at mid-distance from the centre. Combined with the fact that shorter trips, and thus more profitable, concentrate in the centre, double auction scenario’s mean trip distance doubled causing vehicles to travel with at least one customer 60\% less while travelling 15\% more kilometres overall with 40\% fewer customers served. Because the taxis were travelling more, 25\% fewer vehicles were available for assignment on average and revenues under an average \textbaht5 surcharge sank 20\%. When assessing the total emissions, they were raised by 10\%, but thanks to dispersing the fleet onto the suburbs, the emissions in central areas of Bangkok decreased by 20\% to 40\%.

The districts of mid-distances from central area of Bangkok in average experience more than 350 requests per hour and approximately five requests per hour per km2, making the districts areas of high demand. In order to make these longer journeys worthwhile without a loss of revenue, the surcharge of ฿5 could be increased either to a higher fixed point or as a variable linked to the trip distance. This, however, assumes that the customer is willing to pay a high price to travel into the city. In addition, the double auction strategy may be unpopular with customers in the central areas due to the longer waiting time or the decreased availability of taxis for them compared to the rejection scenario. Therefore, the taxi pricing strategy may need to be considered within the wider context of overall transportation supply with public transport modes, so that there are alternative travel options for customers.

To the best of the authors’ knowledge, the double auction pricing mechanism currently isn’t in practice, although recommended by Egan et al \cite{20_Egan_book} and discussed by others \cite{35_ACHEAMPONG}. For the proposed methodology to be implemented, further steps of the algorithm testing and development is required. Double auction balanced-market formation is potentially driven by the vehicle-customer ratio $\eta_{(v/c)}$, which could be deduced from a multi-objective optimization to minimize global emissions and the gap between supply and demand and maximizing profitability. Other parameters such as fleet size or average bid could also be optimized. This trade-off could characterize the limits of fleet efficiency and coverage.

A sensitivity analysis and monte-carlo simulations can be undertaken with more random seeds to initialize the taxi fleet evaluated. 

Introducing taxi-point stations and other methods to calibrate the rejection algorithm to produce the similar supply trends as the probe data \cite{25_iTICFoundation} can help further understand the double auction’s potential to intelligently redistribute the fleet according to demand patterns. Other double auction mechanisms, one-sided auctions, or hybrid scenarios combining double auction and rejection algorithms can also be compared to the developed double auction algorithm.

The potential demand estimation methodology can be complemented with observational surveys or additional inputs to understand the instability added into the market by considering trip rejection, characterised by the break in the ascending trend of success rate and reduction of the supply-demand gap with higher demand levels broken between the 0\% and 5\% in the rejection scenario.

Finally, evaluating the potential shift from private vehicles to public transport resulting from the added service accessibility might draw relevant conclusions on environmental effect, system’s popularity among users in central areas due to the taxi redistribution and the induced demand to other public transport modes. 

% %TODO
% \textcolor{red}{More text on larger implications and transport POLICY is needed}

\section{Conclusions} \label{Sect7:Conclusion}
This paper presented the analysis of a new taxi fare pricing scheme based on double auction and the evaluation of its performance compared to the current taxi landscape with passenger rejections using a bespoke agent-based model of taxis in BMR.

The results show that the taxi pricing and assignment using double auction could be a promising alternative to the current taxi fare scheme in BMR between the city of Bangkok and the neighboring provinces. A double auction fare scheme in conjunction with a new geographical clustering method could increase service accessibility in high-demand yet currently low-supply areas. The newly developed double auction pricing scheme particularly benefits the potential taxi users going from or to mid-distance of 20-40 km away from central part of Bangkok, which currently are perceived as low-profit journeys and rejected by the taxi drivers. The paper acknowledges the future steps of varying surcharges to increase the profit and an adapted clustering method to limit the average waiting time for customers for pickup.

Overall, the paper shows the potential of double auction taxi pricing methodology to boost service accessibility and transport equity to provide taxis to those who otherwise have no access to alternative transport modes to private cars.

\section*{Acknowledgements}
The authors would like to thank UK Research and Innovation funded EPSRC Centre for Doctoral Training in Sustainable Civil Engineering and Lloyd’s Register Foundation, Erasmus program funded by the European Union, and Her Majesty the Queen Sirikit's Scholarship for sponsoring this research. The authors would also like to thank the Intelligent Traffic Information Center Foundation (iTIC), Thailand for providing taxi GPS data. In addition, many thanks to Napameth Phantawesak for providing local knowledge and translation from Thai to English.

\bibliography{main.bbl}

\begin{thebibliography}{10}

\bibitem{1_provinAdmin}
{Department of Provincial Administration}.
\newblock Population of bangkok for year 2563 (2020 on gregorian calendar).
\newblock {\em Population and house statistics report}, 2021.
\newblock Accessed on July 19, 2021.

\bibitem{2_atlas}
Doryane Kermel-Torres.
\newblock Bangkok and the bangkok metropolitan region.
\newblock {\em Atlas of Thailand: Spatial structures and development}, 2004.
\newblock An optional note.

\bibitem{3_IUDA}
{International Urban Development Association}.
\newblock Bangkok metropolitan area, thailand.
\newblock
  \url{https://inta-aivn.org/download/metropolitan-strategies-bangkok/}.
\newblock Accessed on July 20, 2021.

\bibitem{4_Suparee}
T.~Suparee.
\newblock Sustainable urban transport in bangkok.
\newblock
  \url{https://www.unescap.org/sites/default/files/Country\%20Report_Thailand-1_SUTI.pdf}.
\newblock Accessed on March 16,2021.

\bibitem{5_RegistrationTaxi}
{Department of Land Transport}.
\newblock Registration of a taxi car. fares and other regulations.
\newblock \url{https://www.dlt.go.th/th/vehicle-registration/view.php?_did=46},
  2020.
\newblock Accessed on May 19, 2021.

\bibitem{6_Phiboonbanakit}
T.~Phiboonbanakit and T.~Horanont.
\newblock Analyzing bangkok city taxi ride: Reforming fares for profit
  sustainability using big data driven model.
\newblock {\em Journal of Big Data}, 8(1):1--7, 2021.

\bibitem{7_Nimmanunta}
K.~Nimmanunta and T.~Amornpetchkul.
\newblock A clampdown on service refusals by bangkok taxis.
\newblock {\em Asian Journal of Management Cases}, 16(1):38--50, 2019.

\bibitem{8_DeptLand}
{Department of Land Transport}.
\newblock Transport statistics report 2019-2020, 2020.

\bibitem{9_Atsawatheerasathien}
Prapatsorn Atsawatheerasathien and Kunnawee Kanitpong.
\newblock The impact of service zones on passenger-rejection behaviour of
  bangkok taxi drivers.
\newblock {\em Proceedings of the Institution of Civil Engineers - Transport},
  pages 1--10, 2021.
\newblock doi: 10.1680/jtran.19.00022; 13.

\bibitem{10_XE}
{XE}.
\newblock Convert 1 thai baht to us dollar - thb to usd.
\newblock \url{https://www.xe.com/currencyconverter/}.
\newblock Accessed on July 29, 2021.

\bibitem{11_Peungnumsai}
A.~Peungnumsai, H.~Miyazaki, A.~Witayangkurn, and S.~M. Kim.
\newblock The impact of service zones on passenger-rejection behaviour of
  bangkok taxi drivers.
\newblock {\em Proceedings of the Institution of Civil Engineers - Transport},
  pages 1--10, 2020.
\newblock doi: 10.1680/jtran.19.00022; 13.

\bibitem{12_Liftago}
{Liftago}.
\newblock Taxi within a few minutes.
\newblock \url{https://www.liftago.com/personal/taxi/}, 2020.

\bibitem{13_Loke}
Jensen Loke and Ivan Jia.
\newblock An alternative auction model in ride sharing platforms.
\newblock
  \url{https://jensenloke.medium.com/alternative-auction-model-in-ride-sharing-platforms-4d2a0e951cb2},
  12 2017.
\newblock Accessed on May 27, 2021.

\bibitem{14_Karamanis}
Renos Karamanis, Eleftherios Anastasiadis, Panagiotis Angeloudis, and Marc
  Stettler.
\newblock Assignment and pricing of shared rides in ride-sourcing using
  combinatorial double auctions.
\newblock {\em IEEE Transactions on Intelligent Transportation Systems},
  22(9):5648--5659, 2021.

\bibitem{15_Kumar}
Dinesh Kumar, Gaurav Baranwal, Zahid Raza, and Deo~Prakash Vidyarthi.
\newblock A systematic study of double auction mechanisms in cloud computing.
\newblock {\em Journal of Systems and Software}, 125:234--255, 2017.
\newblock ID: 271629.

\bibitem{16_Myerson}
Roger~B Myerson and Mark~A Satterthwaite.
\newblock Efficient mechanisms for bilateral trading.
\newblock {\em Journal of Economic Theory}, 29(2):265--281, 1983.

\bibitem{17_Georgiadis}
George Georgiadis.
\newblock Information economics - module 18: Vcg mechanism.
\newblock
  \url{https://www.kellogg.northwestern.edu/faculty/georgiadis/Teaching/Ec515_Module18.pdf},
  2014.
\newblock Accessed on Feb 14, 2022.

\bibitem{18_Babaioff}
Moshe Babaioff and Noam Nisan.
\newblock Concurrent auctions across the supply chain.
\newblock In {\em Proceedings of the 3rd ACM Conference on Electronic
  Commerce}, EC '01, page 1–10, New York, NY, USA, 2001. Association for
  Computing Machinery.

\bibitem{19_McAfee}
R.~Preston McAfee.
\newblock A dominant strategy double auction.
\newblock {\em Journal of Economic Theory}, 56(2):434--450, 1992.
\newblock ID: 272399.

\bibitem{20_Egan_book}
Malcolm Egan, Martin Schaefer, Michal Jakob, and Nir Oren.
\newblock A double auction mechanism for on-demand transport networks.
\newblock In {\em {PRIMA} 2015: Principles and Practice of Multi-Agent
  Systems}, pages 557--565. Springer International Publishing, 2015.

\bibitem{21_Zhou}
L.~Zhou and H.~Xu.
\newblock An efficient double auction mechanism for on-demand transport
  services in cloud-based mobile commerce.
\newblock In {\em 2017 5th IEEE International Conference on Mobile Cloud
  Computing, Services and Engineering (MobileCloud)}, pages 25--30, Los
  Alamitos, CA, USA, apr 2017. IEEE Computer Society.

\bibitem{22_Egan}
M.~Egan, J.~Drchal, J.~Mrkos, and M.~Jakob.
\newblock Towards data-driven on-demand transport.
\newblock {\em EAI Endorsed Transactions on Industrial Networks and Intelligent
  Systems}, 5:1--10, 2018.

\bibitem{23_Zhang}
Jie Zhang, Ding Wen, and Shuai Zeng.
\newblock A discounted trade reduction mechanism for dynamic ridesharing
  pricing.
\newblock {\em IEEE Transactions on Intelligent Transportation Systems},
  17(6):1586--1595, 2016.

\bibitem{24_OpenStreetMap}
OpenStreetMap.
\newblock Openstreetmap.
\newblock \url{https://www.openstreetmap.org/}, 2021.
\newblock Accessed on July 21, 2022.

\bibitem{25_iTICFoundation}
{Intelligent Traffic Information Center Foundation}.
\newblock itic foundation.
\newblock \url{https://iticfoundation.org/service/open-data-sharing/}, 2021.
\newblock Accessed on January 30, 2021.

\bibitem{26_Siangsuebchart}
Songkorn Siangsuebchart, Sarawut Ninsawat, Apichon Witayangkurn, and Surachet
  Pravinvongvuth.
\newblock Public transport gps probe and rail gate data for assessing the
  pattern of human mobility in the bangkok metropolitan region, thailand.
\newblock {\em Sustainability}, 13(4), 2021.

\bibitem{27_Pueboobpaphan}
Suthatip Pueboobpaphan, Nakorn Indra-Payoong, and Sathaporn Opasanon.
\newblock Experimental analysis of variable surcharge policy of taxi service
  auction.
\newblock {\em Transport Policy}, 76:134--148, 2019.

\bibitem{28_Uttamang}
Pornpan Uttamang.
\newblock {\em Modeling and Analysis of Air Quality in Bangkok Metropolitan
  Region}.
\newblock PhD thesis, North Carolina State University, Raleigh, North Carolina,
  2019.
\newblock Accessed on February 15, 2021.

\bibitem{29_CarDekho}
CarDekho.
\newblock Toyota corolla altis 2008-2013 1.8 g cng on road price, features \&
  specs, images, 2020.

\bibitem{30_Cheewaphongphan}
Penwadee Cheewaphongphan, Agapol Junpen, Savitri Garivait, and Satoru Chatani.
\newblock Emission inventory of on-road transport in bangkok metropolitan
  region (bmr) development during 2007 to 2015 using the gains model.
\newblock {\em Atmosphere}, 8(9), 2017.

\bibitem{31_smit}
R.~Smit, A.~L. Brown, and Y.~C. Chan.
\newblock Do air pollution emissions and fuel consumption models for roadways
  include the effects of congestion in the roadway traffic flow?
\newblock {\em Environmental Modelling \& Software}, 23(10):1262--1270, 2008.
\newblock ID: 271872.

\bibitem{32_Boulter}
Paul Boulter, Ian~S. McCrae, and Tim Barlow.
\newblock A review of instantaneous emission models for road vehicles.
\newblock {\em TRL Published Project Report}, 2007.

\bibitem{33_Zhai}
Zhiqiang Zhai, Ran Tu, Junshi Xu, An~Wang, and Marianne Hatzopoulou.
\newblock Capturing the variability in instantaneous vehicle emissions based on
  field test data.
\newblock {\em Atmosphere}, 11(7), 2020.

\bibitem{34_Gkatzoflias}
Dimitrios Gkatzoflias, Chariton Kouridis, Leonidas Ntziachristos, and Zissis
  Samaras.
\newblock Copert 4: Computer programme to calculate emissions from road
  transport.
\newblock {\em European Environment Agency}, 2006.

\bibitem{35_ACHEAMPONG}
Ransford~A. Acheampong, Alhassan Siiba, Dennis~K. Okyere, and Justice~P.
  Tuffour.
\newblock Mobility-on-demand: An empirical study of internet-based ride-hailing
  adoption factors, travel characteristics and mode substitution effects.
\newblock {\em Transportation Research Part C: Emerging Technologies},
  115:102638, 2020.

\end{thebibliography}

\end{document}